\documentclass[conference]{IEEEtran}

\usepackage{amsmath,amscd,fancyhdr,graphics,pifont}
\usepackage{floatflt,wrapfig, subfigure,epsfig,moreverb,multicol,lscape}
\usepackage{supertabular,tabularx,multirow,amssymb}
\usepackage{theorem,psfrag,afterpage,curves,epic, url}
\usepackage[english]{babel}

\renewcommand{\baselinestretch}{1}

\newcommand{\ignore}[1]{}
\newcommand{\supp}{\operatorname{supp}}

\newcommand{\dmins}{d_{\mathrm{min}}}
\newcommand{\matr}[1]{\mathbf{#1}}
\newcommand{\vect}[1]{\mathbf{#1}}
\newcommand{\code}[1]{\mathcal{#1}}
\newcommand{\cC}{\mathcal{C}}
\newcommand{\set}[1]{\mathcal{#1}}

\newcommand{\GF}[1]{\mathbb{F}_{#1}}
\newcommand{\R}{\mathbb{R}}

\newcommand{\Rp}{\mathbb{R}_{+}}

\newcommand{\tr}{\mathsf{T}}

\newcommand{\codeCQC}[1]{\code{C}_{\mathrm{QC}}^{(r)}}

\newcommand{\defeq}{\triangleq}

\newcommand{\vomega}{\boldsymbol{\omega}}

\newcommand{\onenorm}[1]{\lVert #1 \rVert_1}
\newcommand{\twonorm}[1]{\lVert #1 \rVert_2}

\newcommand{\vc}{\vect{c}}

\newcommand{\Z}{\mathbb{Z}}

\newcommand{\C}{{\mathbb C}}

\renewcommand{\leq}{\leqslant}
\renewcommand{\geq}{\geqslant}

\newtheorem{lemma}{Lemma}
\newtheorem{theorem}[lemma]{Theorem}

\newtheorem{corollary}[lemma]{Corollary}

\theoremstyle{plain}

\newtheorem{PreDefinition}[lemma]{{\textbf{Definition}}}
  \newenvironment{definition}%
    {\begin{PreDefinition}}{\hfill$\square$\end{PreDefinition}}

\theoremstyle{plain}

\newtheorem{PreRemark}[lemma]{{\textbf{Remark}}}
    {\begin{PreRemark}\upshape}{\hfill$\square$\end{PreRemark}}

\newtheorem{PreExample}[lemma]{{\textbf{Example}}}
  \newenvironment{example}%
    {\begin{PreExample}\upshape}{\hfill$\square$\end{PreExample}}

\newcommand{\fch}[2]{\set{#1}(\matr{#2})}

\newcommand{\wps}{w_{\mathrm{p}}}
\newcommand{\wpsAWGNC}{w_{\mathrm{p}}^{\mathrm{AWGNC}}}

\newcommand{\wpsBEC}{w_{\mathrm{p}}^{\mathrm{BEC}}}
\newcommand{\wpsmin}{w_{\mathrm{p}}^{\mathrm{min}}}

\newcommand{\setI}{\set{I}}
\newcommand{\setJ}{\set{J}}

\newcommand{\vtype}{\vect{t}}

\newcommand{\Ftwo}{{{\mathbb F}}_{\!2}}

\begin{document}

%***************************************************************************
% Title
%***************************************************************************

\title{Spectral Graph Analysis of Quasi-Cyclic Codes
%\thanks{%
%   
%    }        
}
\author{
\IEEEauthorblockN{Roxana Smarandache}
\IEEEauthorblockA{Department of Mathematics and Statistics \\
San Diego State University \\
San Diego, CA 92182, USA \\
Email: rsmarand@sciences.sdsu.edu}
\and
\IEEEauthorblockN{Mark F. Flanagan}
\IEEEauthorblockA{Department of Electronic Engineering \\
University College Dublin \\
Belfield, Dublin 4, Ireland \\
Email: mark.flanagan@ieee.org}
}

\maketitle

% \vspace{-7cm}
% \begin{flushright}
%   \texttt{\Statusstring}\\[1cm]
% \end{flushright}
% \vspace{+4cm}

%***************************************************************************
% Main Part
%***************************************************************************

%***************************************************************************
% Abstract
%***************************************************************************
%%%%%%%%%%%%%%%%%%%%%%%%%%%%%%%%%%%%%%%%%%%%%%%%%%%%
\begin{abstract}

  In this paper we analyze the bound on the additive white Gaussian
  noise channel (AWGNC) pseudo-weight of a $(c,d)$-regular linear
  block code based on the two largest values $\lambda_1 >\lambda_2$ of
  the eigenvalues of $\matr{H}^{\tr}\matr{H}$: $\wpsmin(\matr{H})\geq
  n\frac{2c-\lambda_2}{\lambda_1-\lambda_2}$.  %
  \cite{Vontobel:Koetter:04:1}.  In particular, we analyze
  $(c,d)$-regular quasi-cyclic (QC) codes of length $rL$ described by
  $J\times L$ block parity-check matrices with circulant block entries
  of size $r\times r$. We proceed by showing how the problem of
  computing the eigenvalues of the $rL\times rL$ matrix
  $\matr{H}^{\tr}\matr{H}$ can be reduced to the problem of computing
  eigenvalues for $r$ matrices of size $L \times L$. We also give a necessary
  condition for the bound to be attained for a circulant matrix
  $\matr{H}$ and show a few classes of cyclic codes satisfying this
  criterion.
\end{abstract}
  
%***************************************************************************
% Keywords
%***************************************************************************

%%%%%%%%%%%%%%%%%%%%%%%%%%
\section{Introduction}
\label{sec:introduction:1}

Low-density parity-check (LDPC) codes offer excellent tradeoffs between performance and complexity for error correction in communication systems. Quasi-cyclic (QC) LDPC codes in particular have proved extremely attractive due to their implementation advantages, both in encoding and decoding \cite{Li_QC_encoders:06, Dai:08, Mansour:07}. Many analyses of QC-LDPC codes have been carried out based on optimization of parameters such as the minimum Hamming distance of the code or the girth of the Tanner graph. However, it has been shown that an excellent first-order measure of performance over the AWGNC is the minimum \emph{pseudo-weight} of the code \cite{Wiberg}. So far, few results exist in the literature on the minimum pseudo-weight of QC-LDPC and related codes. 

Spectral graph analysis was used in~\cite{Tanner:01:1}, and more
recently, in~\cite{Vontobel:Koetter:04:1}, to obtain bounds on the
minimum Hamming weight, and minimum AWGNC pseudo-weight, respectively, of a length-$n$
$(c,d)$-regular code $\cC$ over the binary field $\Ftwo$:
  $$\dmins\geq\wpsmin(\matr{H}) \geq n  \frac{2c -
    \lambda_2}{\lambda_1 - \lambda_2}; \dmins \geq n\frac{2}{d}
  \frac{2c +d - 2 - \lambda_2}{\lambda_1 - \lambda_2},$$ with
  $\lambda_1 > \cdots > \lambda_s$ being the distinct ordered
  eigenvalues of $\matr{H}^{\tr} \matr{H}\in \R^{n\times n}$ (where $\matr{H}$ is viewed as
  a matrix in $ \R^{m\times n}$). These bounds are, for most codes,
  loose.  However, in particular cases, like the projective geometry
  codes \cite{Vontobel:Smarandache:05:1, Smarandache:Vontobel:07,Kou:Lin:Fossorier:01:1}, they are attained.
  A current problem with these bounds is that for most LDPC codes, it is not practical to evaluate the eigenvalues $\lambda_1,\lambda_2$ due to the size of the matrix $\matr{H}^{\tr} \matr{H}$. 
  
  In this paper we show how to compute the AWGN pseudo-weight lower bound for quasi-cyclic
  (QC) and related codes by utilizing the $\mathcal A$-submodule
  structure of quasi-cyclic codes, ${\cal A}= \R[X]/(X^r-1)$ \cite{Lally:Fitzpatrick:01, Ling:Sole:01, Ling:Sole:03}. 
  In particular, we begin by showing how the
  polynomial parity-check matrix that describes a cyclic code can be used
  to compute the required eigenvalues, and then generalize this approach to compute the required eigenvalues for QC codes. 
  We also define the class of ``nested circulant'' matrices, and show that these have eigenvalues
  which are given by evaluating a multivariate associated
  polynomial at points whose coordinates are particular roots of
  unity. Finally, we give a necessary 
  condition for the pseudo-weight lower bound to be attained when $\matr{H}$ is circulant
  and show a few classes of cyclic codes satisfying this criterion.

%***************************************************************************

\section{Basic Notation and Definitions}
\label{sec:notation:1}

All codes in this paper will be binary linear codes of a certain
length $n$ specified through a (scalar) parity-check matrix
$\matr{H}=(h_{j,i}) \in \GF{2}^{m \times n}$ as the set of all
vectors $\vc \in \Ftwo^n$ \ such that $\matr{H} \cdot \vc^\tr =
\vect{0}^\tr$, where ${}^\tr$ denotes transposition. The minimum
Hamming distance of a code $\code{C}$ will be denoted by $\dmins(\code{C})$.  The
fundamental cone $\fch{K}{H}$ of $\matr{H}$ is the set of all vectors
$\vomega \in \R^n$ that satisfy
  \begin{alignat}{2}
    \omega_i
      &\geq 0 
      \ 
      &&\text{for all $i \in \setI(\matr{H})$} \; , 
          \label{eq:fund:cone:def:1} \\
    \omega_i
      &\leq
          \sum_{i' \in \setI_j(\matr{H}) \setminus i} \!\!
            \omega_{i'}
      \ 
      &&\text{for all $j \in \setJ(\matr{H})$, \ 
               $i \in \setI_j(\matr{H})$} \; ,
          \label{eq:fund:cone:def:2}
  \end{alignat}
  where $\setJ(\matr{H})$ and $\setI(\matr{H})$ denote the sets of row
  and column indices of $\matr{H}$ respectively,  and $\setI_j(\matr{H}) \defeq \{ i \in
  \setI \ | \ h_{j,i} = 1 \}$ for each $j \in \setJ(\matr{H})$.  A vector $\vomega \in \fch{K}{H}$ is called a \emph{pseudo-codeword}.  The AWGNC \emph{pseudo-weight} of a
  pseudo-codeword $\vomega$ is defined to be $\wps(\vomega) =
  \wpsAWGNC(\vomega) \defeq \lVert \vomega \rVert_1^2 / \lVert \vomega
  \rVert_2^2$. (For a motivation of these definitions,
  see~\cite{Vontobel:Koetter:05:1:subm,
    Koetter:Li:Vontobel:Walker:07:1}).  The minimum of the AWGNC
  pseudo-weight over all nonzero pseudo-codewords is called the minimum
  AWGNC pseudo-weight and is denoted by $\wpsmin(\matr{H})$.

For any integer $s \ge 1$, let $R_s = \{ \exp(\imath 2 \pi r / s) \; : \; 0 \le r < s \}$
denote the set of complex $s$-th roots of unity, and let $R_s^{-} =
R_s\backslash \{ 1 \}$. The symbol ${}^*$ denotes complex conjugation. Also, an $r \times r$ circulant matrix $\matr{B}$, whose entries are square $L \times L$ matrices, will be called an \emph{$L$-block circulant} matrix; we shall denote this by
\[
\matr{B} = \mathrm{circ}(\matr{b}_0, \matr{b}_1, \cdots, \matr{b}_{r-1})
\] 
where the (square $L \times L$ matrix) entries in the first column of $\matr{B}$ are $\matr{b}_0$, $\matr{b}_1$, ... , $\matr{b}_{r-1}$ respectively. 

  Finally, $\Z$, $\R$, $\C$, and $\Ftwo$ will be the ring of integers,
  the field of real numbers, the complex field, and the finite field
  of size $2$, respectively. For a positive integer $L$, $[L]$ will
  denote the set of nonnegative integers smaller than $L$:
  $[L]=\{0,1,\ldots, L-1\}$.

  \section{Computing the Eigenvalues of $\matr{H}^{\tr}\matr{H}$ 
for a QC Code}
\label{sec:eigenvalues}

In this section we will show that the polynomial representation of a
QC code will prove very helpful in computing the eigenvalues of the
large matrix $\matr{H}^{\tr}\matr{H}$, easing in this way the computation
of the lower bound
\begin{align}
\label{lowerbound} \dmins&\geq\wpsmin(\matr{H}) \geq n  \frac{2c -
    \lambda_2}{\lambda_1 - \lambda_2}.
\end{align}
This section is organized in three subsections. In Sec.
\ref{subsec:circulantmatrix} and \ref{subsec:QCcodes} we provide
some background on circulant matrices and QC codes. Section
\ref{subsec:eigenvaluesQC} will contain the main result on the
eigenvalues of $\matr{H}^{\tr}\matr{H}$, where $\matr{H}$ is the
parity-check matrix of a QC code.
\subsection{ Eigenvalues of a Circulant Matrix} 
\label{subsec:circulantmatrix}

The eigenvalues of a square circulant matrix are well known
\cite{MacWilliams:Sloane:98}.  If $\matr{B}\in \C^{n\times n}$ is a
circulant matrix and $w(X)= b_0+b_1X+\ldots +b_{n-1}X^{n-1}$ its
(column) associated polynomial, then the eigenvalues of $\matr{B}$ are
given by this polynomial's evaluation at the complex $n$-th roots of unity, i.e. $w(x)$ for all $x \in R_n$. 

The following gives a proof of this result based on the polynomial representation of a circulant
matrix. It may be seen as a special case of the method we present later for QC codes.

Let  $\lambda$ be an eigenvalue of $\matr{B}$. Then there exists a nonzero
vector $\vect{v}=(v_0, \ldots,
 v_{n-1})^\tr \in \C^{n}$ such that
\begin{align*}
&\matr{B}\vect{v}=\lambda\vect{v}.
 \end{align*}
 In polynomial form, this equation is equivalent to (here $v(X) = v_0+v_1X+\ldots +v_{n-1}X^{n-1}$):
 \begin{align*}
   &w(X)v(X)=\lambda v(X) \mod (X^n-1) {~\rm iff}
   \\& X^n-1~|~w(X)v(X)-\lambda v(X) {~\rm in~} \C{~\rm iff}\\
   &w(x)v(x)=\lambda v(x), \forall x \in R_n{~\rm iff}\\
   &(w(x)-\lambda)v(x)=0, \forall x\in R_n \; .\\
%&\lambda=c(\rho^i)v(\rho^i)/v(\rho^i)=c(\rho^i)
\end{align*}
For each $x\in R_n$, $\lambda= w(x)$ is a solution of the above
equation, and therefore it is an eigenvalue for the matrix $\matr{B}$.
There are $n$ such solutions, therefore, these are all possible
eigenvalues of $\matr{B}$.

In the next theorem we will consider an \emph{$L$-block circulant} matrix instead of a circulant matrix. This theorem may be found in \cite{Tee:05}; 
we provide here an alternative proof based on the polynomial representation.
\begin{theorem}\label{theorem2} 
  Let $\matr{B}=\mathrm{circ}(\matr{b}_0, \matr{b}_1, \cdots, \matr{b}_{r-1})\in \C^{rL\times rL}$ be an
  $L$-block circulant matrix.  Let $\matr{W}(X)= \matr{b}_0+\matr{b}_1X+\ldots
  +\matr{b}_{r-1}X^{r-1}$ its (column) associated matrix polynomial.
  Then the eigenvalues of $\matr{B}$ are given by the union of the
  eigenvalues of the $L\times L$ matrices $\matr{W}(x)$, for all $x\in
  R_r$.
\end{theorem}   
 
\IEEEproof The proof follows the reasoning in the theorem above.

Let $\lambda$ be an eigenvalue of $\matr{B}$. Then there exists a nonzero
vector $\vect{v}\defeq(v_0, \ldots, v_{rL-1})^\tr\in \C^{rL}$ such
that
\begin{align}\label{eigenvalue}
  &\matr{B}\vect{v}=\lambda\vect{v}.
 \end{align}
 Let $\vect{p}(X)\in \C^{L}[X]$ given by $\vect{p}(X)=(v_0, \ldots,
 v_{L-1})^\tr+(v_L, \ldots, v_{2L-1})^\tr X+\ldots+ (v_{r(L-1)},
 \ldots, v_{rL-1})^\tr X^{r-1}.$ In polynomial form,
 equation~\eqref{eigenvalue} is equivalent to:
 \begin{align*}
   &\matr{B}(X) \vect{p}(X)=\lambda \vect{p}(X) \mod (X^r-1) {~\rm iff}
   \\& X^r-1~|~\matr{B}(X)\vect{p}(X)-\lambda \vect{p}(X) {~\rm in~} \C {~\rm iff}\\
   &\matr{B}(x)\vect{p}(x)=\lambda \vect{p}(x), \forall x\in R_r.
\end{align*}
The last equation is the equation for the eigenvalues of the matrix
$\matr{B}(x)$. Each such matrix has $L$ eigenvalues, counting
multiplicities, and there are $r$ distinct complex numbers in $R_r$; this accounts for the total number $rL$ of eigenvalues of
$\matr{B}$.  The eigenvectors can also be deduced from the above.

% NOTE: Need to make sure we handle multiplicities correctly. Does Tee's
% result include multiplicities? If not then it makes a good case for
% this result.

\subsection{Definition and Properties of QC Codes}  
\label{subsec:QCcodes}
A linear QC-LDPC code $\code{C}_{\rm QC}\defeq\codeCQC{r}$ of length
$n = rL$ can be described by an $rJ \times rL$ (scalar) parity-check
matrix $\matr{\bar H}_{\rm QC}^{(r)}\defeq\matr{\bar H}$ that is formed by a $J
\times L$ array of $r \times r$ circulant matrices.
\begin{align}
\matr{\bar H}=\left[
\begin{array}{cccc}
\matr{P}_{1,1}   &       \matr{P}_{1,2}   &       \ldots  &       \matr{P}_{1,L} \\
\matr{P}_{2,1}   &       \matr{P}_{2,2}  &       \ldots   &       \matr{P}_{2,L}  \\
\vdots &\vdots &\ldots &\vdots \\
\matr{P}_{J,1}  &       \matr{P}_{J, 2}  &    \ldots   &       \matr{P}_{J, L} 
\end{array}
 \right], 
\end{align}
where the entries $\matr{P}_{i,j}$ are $r\times r$ circulant matrices.
Clearly, by choosing these circulant matrices to be low-density, the
parity-check matrix will also be low-density.

With the help of the well-known isomorphism between the ring of
$r\times r$ circulant matrices and the ring of polynomials modulo $X^r
- 1$, to each matrix $\matr{P}_{i,j}$ we can associate a polynomial
$p_{i,j}(X)$, and thus a QC-LDPC code can equivalently be described by
a polynomial parity-check matrix $\matr{P}(X)$ of size $J \times L$,
with polynomial operations performed modulo $X^r-1$:
\begin{align}
\matr{P}(X)=\left[
\begin{array}{cccc}
p_{1,1}(X)   &       p_{1,2}(X)  &       \ldots  &       p_{1,L}(X) \\
p_{2,1}(X)   &       p_{2,2}(X)  &       \ldots   &       p_{2,L}(X)  \\
\vdots &\vdots &\ldots &\vdots \\
p_{J,1}(X)  &       p_{J, 2}(X)  &    \ldots   &       p_{J, L}(X) 
\end{array}
 \right].
\end{align}
 
By permuting the rows and columns of the scalar parity-check matrix
$\matr{\bar H}$,\footnote{i.e., by taking the first row in the first block
  of $r$ rows, the first row in the second block of $r$ rows, etc.,
  then the second row in the first block, the second row in the second
  block, etc., and similarly for the columns.} we obtain an equivalent
parity-check matrix representation $\matr{H}$ for the QC code
$\codeCQC{r}$,

\begin{align}\matr{H}
    &\defeq
       \begin{bmatrix}
         \matr{H}_0      & \matr{H}_{r-1}      & \cdots 
                         & \matr{H}_1 \\ 
         \matr{H}_1      & \matr{H}_0          & \cdots   
                         & \matr{H}_2 \\ 
         \vdots          & \vdots              & \ddots           
                         & \vdots \\ 
                  \matr{H}_{r-1}   & \matr{H}_{r-2} & \cdots 
                         & \matr{H}_0
  \end{bmatrix}.
\label{eq:matrix_1_bijection}
\end{align}
where $\matr{H}_0, \matr{H}_1, \ldots, \matr{H}_{r-1}$ are scalar $J
\times L$ matrices.  The connection between the two representations is
\begin{align}
\matr{H}_0 + \matr{H}_1 X + \cdots + \matr{H}_{r-1} X^{r-1}=\matr{P}(X).
\label{eq:matrix_2_bijection}
\end{align}

\subsection{The Eigenvalues of the Matrix $\matr{H}^\tr\cdot \matr{H}$ of a QC Code}
\label{subsec:eigenvaluesQC}

Note that for a fixed value of $r \ge 1$, (\ref{eq:matrix_2_bijection}) provides a simple bijective correspondence between the set of polynomial matrices $\matr{P}(X) \in (\R[X]/(X^r-1))^{J \times L}$ and the set of parity-check matrices of the form (\ref{eq:matrix_1_bijection}). Furthermore, the product of two such polynomial matrices, where defined, yields another which corresponds via this bijection with the product of the corresponding parity-check matrices in the form (\ref{eq:matrix_1_bijection}). Also note that transposition of a polynomial matrix in the form (\ref{eq:matrix_2_bijection}) corresponds to transposition of the corresponding parity-check matrix in the form (\ref{eq:matrix_1_bijection}), under this bijection.

%As mentioned above, the set of $r\times r$ circulant matrices is a
%ring isomorphic to the ring of polynomials modulo $x^r - 1$.
%Therefore multiplying two circulant matrices gives a circulant matrix
%and the polynomial associated to the product is equal to the product
%of the two corresponding polynomials. A similar statement may be made
%about the ring of block circulant matrices with $r\times r$ circulant matrix entries, isomorphic to the corresponding matrix polynomial ring (note also that the transpose of such a matrix yields 
%another such matrix). 
It follows that $\matr{H}^\tr\cdot\matr{H}$ is an $L$-block circulant matrix; applying Theorem~\ref{theorem2} to this matrix yields the following corollary.

\begin{corollary}

  The eigenvalues of $\matr{H}^\tr\cdot\matr{H}$ are given
  by the union of the eigenvalues of the $L\times L$ matrices
  $\matr{P}^\tr(x^*)\cdot \matr{P}(x),$ for $x\in R_r$.
\label{cor:QC_codes}
\end{corollary} 
\IEEEproof We apply Theorem~\ref{theorem2} to the $L$-block circulant
matrix $\matr{H}^\tr\cdot\matr{H}\defeq \mathrm{circ}(\matr{b}_0, \matr{b}_1, \cdots, \matr{b}_{r-1})\in \C^{rL\times rL}$ and form the matrix $\matr{W}(X)= \matr{b}_0+\matr{b}_1X+\ldots
+\matr{b}_{r-1}X^{r-1}$.  This is equal to the product of the two
matrix polynomials of $\matr{H}^\tr$ and $\matr{H}$,
which are $\matr{H}_0^{\tr} + \matr{H}_{r-1}^{\tr} X +
\cdots + \matr{H}_{1}^{\tr} X^{r-1} = X^r\matr{P}^\tr(1/X)$ and $\matr{H}_0 + \matr{H}_1 X +
\cdots + \matr{H}_{r-1} X^{r-1} = \matr{P}(X)$, respectively. Therefore
$\matr{W}(X)=(X^r\matr{P}^\tr(1/X))\cdot\matr{P}(X)$ and so the eigenvalues of $\matr{H}^\tr\cdot\matr{H}$ are the eigenvalues of $\matr{P}^\tr(1/x)\cdot
\matr{P}(x)$, for all $x\in R_r$; these are then equal to the eigenvalues of $\matr{P}^\tr(x^*)\cdot
\matr{P}(x)$, for all $x\in R_r$ (as $x^*=1/x$ for all such $x$).

\begin{example}\label{Tannerexample}
  Let $r=31$ and consider the $(3,5)$-regular QC-LDPC code given by the
  scalar $93 \times 155$ matrix\footnote{Here $\matr{I}_\ell$ denotes the $31 \times 31$ identity matrix with rows
shifted cyclically to the left by $\ell$ positions.}
\begin{align*}
  {\matr{\bar H}} &= \begin{bmatrix}
    \matr{I}_1   & \matr{I}_2    & \matr{I}_4    & \matr{I}_8 & \matr{I}_{16}\\
    \matr{I}_5   & \matr{I}_{10} & \matr{I}_{20} & \matr{I}_9 & \matr{I}_{18}\\
    \matr{I}_{25}& \matr{I}_{19} & \matr{I}_7 & \matr{I}_{14}&
    \matr{I}_{28}
         \end{bmatrix}.
   \end{align*}
   The polynomial parity-check matrix $\matr{P}(X)\in
   (\R[X]/(X^r-1))^{3 \times 5}$ is
  \begin{align*}
    \matr{P}(X)
       &= \begin{bmatrix}
           X      & X^2    & X^4    & X^8 & X^{16}\\
           X^5    & X^{10} & X^{20} & X^9 & X^{18}\\
           X^{25} & X^{19} & X^7    & X^{14}& X^{28}
         \end{bmatrix} \; .
  \end{align*}
  This code is the famous $(3,5)$-regular QC-LDPC code of length $155$
  presented in~\cite{Tanner:Sridhara:Fuja:01:1}. Note that the code
  parameters are $[155, 64, 20]$.  The corresponding matrix
  $\matr{H}$ in the form (\ref{eq:matrix_1_bijection}) is a $31\times 31 $ matrix with block entries
  $\matr{H}_i$, $i\in [31]$ obtained by decomposing $\matr{P}(X)$
  according to the powers of $X$:
  \begin{align}
\matr{P}(X)=\matr{H}_0 + \matr{H}_1 X + \cdots + \matr{H}_{30} X^{30}.
\end{align}
Obviously only $15$ matrices among the $\matr{H}_i$ are nonzero, and all
of these contain only one $1$, the other entries being zero.

The matrix $\matr{H}^\tr\cdot\matr{H}$ is a $5$-block circulant matrix. Corollary \ref{cor:QC_codes} above tells us
that in order to compute its eigenvalues, we need to form the matrices
$\matr{P}^\tr(\rho^{-i})\cdot \matr{P}(\rho^i)$, for all $i\in [31]$ (here $\rho$ denotes a primitive complex $31$-th root of unity). We
have that
\begin{align*}
  \matr{P}^\tr(1/x) &= \begin{bmatrix}
    x^{30}      & x^{29}    & x^{27}    & x^{23} & x^{15}\\
    x^{26}    & x^{21} & x^{11} & x^{22} & x^{13}\\
    x^{6} & x^{12} & x^{24} & x^{17}& x^{3}
         \end{bmatrix}^\tr \;
  \end{align*}
and
%\begin{align*}
%  &  \matr{P}^\tr(1/x)\cdot \matr{P}(x)=
%  \begin{bmatrix}
%    3  &a & \overline{e} & c  &\overline{e} \\
%    \overline{a}& 3& b &\overline{a}& d\\
%    e& \overline{b}& 3 &c &\overline{b}\\
%    \overline{c}& a& \overline{c}& 3 &d\\
%    e &\overline{d}& b& \overline{d} &3\end{bmatrix} \;,
%\end{align*}
\begin{align*}
  &  \matr{P}^\tr(1/x)\cdot \matr{P}(x)=
  \begin{bmatrix}
    3  & a & e^* & c  & e^* \\
    a^* & 3 & b & a^* & d \\
    e & b^* & 3 & c & b^* \\
    c^* & a & c^* & 3 & d \\
    e & d^* & b & d^* & 3 \end{bmatrix} \;,
\end{align*}
for all $x \in R_{31}$, where 
%\begin{align*}
%  a&=\rho + \rho^5 + \rho^{25}; b = \rho^2 + \rho^{10} + \rho^{19};
%  c = \rho^4 + \rho^{20} + \rho^7;\\
%  d& =\rho^8 + \rho^9 + \rho^{14}; e = \rho^{16} + \rho^{18} +
%  \rho^{28}.
%\end{align*}
\begin{align*}
  a&=x + x^5 + x^{25}; b = x^2 + x^{10} + x^{19};
  c = x^4 + x^7 + x^{20};\\
  d& =x^8 + x^9 + x^{14}; e = x^{16} + x^{18} +
  x^{28}.
\end{align*}
Obviously for $i\in [31]$, each matrix $\matr{P}^\tr(\rho^{-i})\cdot
\matr{P}(\rho^i)$ is Hermitian (in fact nonnegative definite), hence each has $5$ real nonnegative eigenvalues,
giving a total of $31\cdot 5=155$ nonnegative eigenvalues for $\matr{H}^\tr\cdot\matr{H}$.

We obtain that for each $i \in [31], i\neq 0$,
the associated polynomial of $\matr{P}^\tr(\rho^{-i})\cdot
\matr{P}(\rho^i)$ may be written as (using $\rho^{31} = 1$)
\begin{eqnarray*}
u(\lambda) & = & \lambda^2(\lambda^3 - 15 \lambda^2 + 62 \lambda - 62) \\
& = & \lambda^2(\lambda-\lambda_2)(\lambda-\lambda_3)(\lambda-\lambda_4)
\end{eqnarray*}
where $\lambda_2 = 8.6801$, $\lambda_3 = 4.8459$ and $\lambda_4 =
1.4740$. Also, for $i=0$ the associated polynomial of
$\matr{P}^\tr(\rho^{-i})\cdot \matr{P}(\rho^i)$ may be written as
$u(\lambda) = \lambda^4(\lambda-\lambda_1)$ where $\lambda_1 = 15$.
This yields the nonzero eigenvalues of $\matr{H}^\tr\cdot\matr{H}$ as $\{ \lambda_1, \lambda_2, \lambda_3,
\lambda_4 \}$ with multiplicities $1$, $30$, $30$ and $30$
respectively.
\end{example} 
 
%These results are based on the paper by Tee \cite{Tee:05}.

\section{Eigenvalues of Nested Circulant Matrices}
\label{sec:nestedcirculanteigenvalues}
In this section we define the class of \emph{nested circulant} matrices,
and show that they have eigenvalues which are given by evaluating a
multivariate associated polynomial at points whose coordinates are
particular roots of unity. 

\begin{theorem}\label{theorem:nested_2} 
  Let $\matr{B}=\mathrm{circ}(\matr{b}_0, \matr{b}_1, \cdots, \matr{b}_{r-1})\in \C^{rL\times rL}$ be an $L$-block
  circulant matrix. Suppose that each subblock $\matr{b}_i$,
  $i \in [r]$, is also circulant, with associated polynomial
  $p^{(i)}(X) = \sum_{j=0}^{L-1} b_{i,j} X^j$. Define the
  associated polynomial of $\matr{B}$ by
\[ 
q(X,Y) = \sum_{i=0}^{r-1} \sum_{j=0}^{L-1} b_{i,j} X^i Y^j \; .
\] 
Then the set of eigenvalues of $\matr{B}$ is given by 
\[
\{ q(x,y) \; : \; x \in R_r, y \in R_L \} \; .
\]
\end{theorem}   

\IEEEproof For each $j \in [L]$ define $u^{(j)}(X) =
\sum_{i=0}^{r-1} b_{i,j} X^i$. By Theorem~\ref{theorem2}, the
eigenvalues of $\matr{B}$ are equal to those of the matrices given by
$\matr{W}(x)$ for $x \in R_r$; each of these is circulant with
associated polynomial (in $Y$) given by
\[
\sum_{j=0}^{L-1} u^{(j)}(x) Y^j = q(x,Y) \; .
\]
Thus the eigenvalues of each $\matr{W}(x)$ are equal to $q(x,y)$ for
$y \in R_L$, and the result follows.

We next define what is meant by a \emph{nested circulant} matrix.
%\vspace{-8mm}
\begin{definition}\label{m_nested_circulant}
Let $m \ge 1$ and let $i_t$ be a positive integer for each $t=1,2,\cdots,m$. Also let $\matr{B} =
  \mathrm{circ}(\matr{b}_0, \matr{b}_1, \cdots, \matr{b}_{i_1-1})$ be
  a block-circulant matrix such that for every $t=1,2,\cdots,m-1$,
  $j_t \in [i_t]$ 
\begin{align*} &\matr{b}_{j_1,j_2,\cdots,j_{t}}=\\ &
  \mathrm{circ}(\matr{b}_{j_1,j_2,\cdots,j_{t},0},
  \matr{b}_{j_1,j_2,\cdots,j_{t},1}, \cdots,
  \matr{b}_{j_1,j_2,\cdots,j_{t},i_{t+1}-1})\end{align*}
is also block-circulant,
  and that $\matr{b}_{j_1,j_2,\cdots,j_{m}} = b_{j_1,j_2,\cdots,j_{m}}$ are
  scalars. Then $\matr{B}$ is said to be an $m$-nested circulant
  matrix (with dimension $n = \prod_{t=1}^{m} i_t$). The associated polynomial of $\matr{B}$ is defined by
\begin{equation}
  q(X_1,X_2,\cdots,X_{m}) = \sum_{j_1=0}^{i_1-1} \sum_{j_2=0}^{i_2-1} 
\cdots \sum_{j_{m}=0}^{i_{m}-1} b_{j_1,j_2,\cdots,j_{m}} \prod_{t=1}^{m} X_{t}^{j_t}   
\label{eq:definition_char_poly} 
\end{equation}
\end{definition}
Note that the $1$-nested circulants are precisely the circulant
matrices, and that the $2$-nested circulants are precisely the 
$i_2$-block-circulant matrices with circulant subblocks. Also note that the
associated polynomial $q(X_1,X_2,\cdots,X_{m})$ provides a succinct
description of the matrix $\matr{B}$.

A straightforward generalization of Theorem \ref{theorem:nested_2} is
as follows.
%\vspace{-8mm}
\begin{theorem}\label{theorem:nested_m} 
  Let $\matr{B}$ be an $m$-nested circulant matrix with associated
  polynomial $q(X_1,X_2,\cdots,X_{m})$ given by
  (\ref{eq:definition_char_poly}) above.  Then the set of eigenvalues
  of $\matr{B}$ is given by
\[
\{ q(x_1,x_2,\cdots,x_{m}) \; : \; x_t \in R_{i_t} \quad \forall t = 1,2,\cdots,m \}
\]
\end{theorem}   

\IEEEproof The proof uses induction, and follows the lines of the
proof of Theorem \ref{theorem:nested_2} in a rather straightforward
manner.

\begin{example}\label{fully_nested_circulant}
  Here we take an example of an $3$-nested circulant (i.e. $m=3$),
  where $i_t=2$ for $t=1,2,3$. The eigenvalues of
\[
\matr{B} = \begin{bmatrix} 
0 & 1 & 0 & 0 & 0 & 1 & 1 & 1 \\
1 & 0 & 0 & 0 & 1 & 0 & 1 & 1 \\
0 & 0 & 0 & 1 & 1 & 1 & 0 & 1\\
0 & 0 & 1 & 0 & 1 & 1 & 1 & 0 \\
0 & 1 & 1 & 1 & 0 & 1 & 0 & 0 \\
1 & 0 & 1 & 1 & 1 & 0 & 0 & 0 \\
1 & 1 & 0 & 1 & 0 & 0 & 0 & 1\\
1 & 1 & 1 & 0 & 0 & 0 & 1 & 0\end{bmatrix}
\]
are equal to the eigenvalues of 
\[
\matr{B'} = \begin{bmatrix} 
0 & 1+x & x & x \\
1+x & 0 & x & x \\
x & x & 0 & 1+x \\
x & x & 1+x & 0\end{bmatrix}
\]
for $x \in \{-1,1\}$, which are equal to the eigenvalues of 
\[
\matr{B''} = \begin{bmatrix} 
xy & 1+x+xy \\
1+x+xy & xy\end{bmatrix}
\]
for $x \in \{-1,1\}$ and $y \in \{-1,1\}$. Finally, these are equal to the set 
\[
\{ q(x,y,z) \; : \; x,y,z \in \{-1,1\} \}
\]
where the associated polynomial of $\matr{B}$ is $q(x,y,z) = xy +
z(1+x+xy)$. In this example $b_{0,0,0} = 0$, $b_{0,0,1} = 1$,
$b_{0,1,0} = 0$, $b_{0,1,1} = 0$, $b_{1,0,0} = 0$, $b_{1,0,1} = 1$,
$b_{1,1,0} = 1$, $b_{1,1,1} = 1$; these may be easily obtained by
matching the elements of the first column of $\matr{B}$ with the binary
expansion of the corresponding row position.

This example may be generalized to the case where $n = 2^m$ and the
circulant is $m$-nested; the eigenvalues are real. % The resulting
%matrices look a little like Hadamard matrices. 
Note that the choice of the first column in $\matr{B}$ determines which
terms in $\{ 1,x,y,z,xy,yz,zx,xyz \}$ are included in the
associated polynomial, and hence controls the eigenvalues of
$\matr{B}$.
\end{example}

\begin{theorem}\label{theorem:nested_H_nested_B}
  If $\matr{H}$ is an $m$-nested circulant matrix, then $\matr{B} =
  \matr{H}^{\tr} \matr{H}$ is an $m$-nested circulant matrix. 
\end{theorem}

\IEEEproof It is straightforward to prove the stronger result that if $\matr{A}$ and $\matr{B}$ are $m$-nested circulants with specified nested dimensions, then $\matr{A}^{\tr} \matr{B}$ is also $m$-nested circulant, with the same nested dimensions. The proof proceeds by induction on $m$. The base case $m=1$ is straightforward. Next, let $\matr{A}$ be block-circulant with block entries in the first column equal to some $(m-1)$-nested circulants $\matr{A}_i$, and let $\matr{B}$ be block-circulant with block entries in the first column equal to some $(m-1)$-nested circulants
$\matr{B}_j$. The matrix $\matr{A}^{\tr} \matr{B}$ is then block-circulant, and each block entry is a sum of matrices of the form $\matr{A}_i^{\tr} \matr{B}_j$. By the principle of induction, each
of these matrices is an $(m-1)$-nested circulant, and it is easy to show that a sum of $t$-nested circulants (of the same nested dimensions) is another $t$-nested circulant (with these nested dimensions).
% \section{Improving the Tanner and Vontobel's bounds on the minimum distance 
%and pseudo-weight} 
% In this section we have to understand how this above splitting can
% help in improving the bound. See the proof and understand it. See how
% we can prove it using this splitting... etc.
% This is where main results come in.

\section{Conditions for the Pseudo-Weight Lower Bound to Hold with Equality}
\label{sec:conditions}  

It is straightforward to show that a necessary condition for the bound of \cite{KV-lower-bounds}
to hold with equality is that the eigenvalues of $\matr{B} =
\matr{H}^{\tr} \matr{H} \in \R^{n\times n}$ are $\lambda_1$ with multiplicity $1$ and
$\lambda_2 < \lambda_1$ with multiplicity $n-1$.

If $\matr{H}$ is circulant with (row) associated polynomial $w(X)$ of degree $k \le n$, the eigenvalues of $\matr{B}$
are precisely $\{ \left| w(x) \right|^2 \; : \; x \in R_n \}$;
therefore the largest eigenvalue of $\matr{B}$ is $\lambda_1 = \left|
  w(1) \right|^2 = d^2$ where $d$ is the number of nonzero
coefficients in $w(X)$ (noting that $\left| w(1) \right|^2 > \left|
  w(x) \right|^2$ for all $x \in R_n^{-}$). Let $\tilde{w}(X) = X^k w(1/X)$ denote the \emph{reciprocal
polynomial} of $w(X)$ which is obtained by reversing the order of
coefficients in $w(X)$. Now assume that the bound of
\cite{KV-lower-bounds} holds with equality. Then we must have
\[
\left| w(x) \right|^2 = w(x)w^{*}(x) = \lambda_2 \quad \forall \: x \in R_n^{-}
\] 
for some positive real number $\lambda_2$, i.e.
\[
w(x)w(1/x) = \lambda_2 \quad \forall \: x \in R_n^{-} \; .
\]
This is equivalent to
\[
w(x)\tilde{w}(x) = \lambda_2 x^k \quad \forall \: x \in R_n^{-}
\]
Thus $R_n^{-}$ is a subset of the roots of the polynomial
$w(X)\tilde{w}(X) - \lambda_2 X^k$, and so
\begin{equation}
w(X)\tilde{w}(X) - \lambda_2 X^k = (1+X+X^2+\cdots +X^{n-1}) r(X)
\label{eq:equality_in_bound_circ}
\end{equation}
where $r(X)$ is a polynomial of degree $2k-n+1 \ge 0$ with integer
coefficients. In the following we give details of this condition for
some codes which attain the bound of \cite{KV-lower-bounds} with
equality.
\vspace{-2mm}
\begin{example}\label{EG22example}
  The $\mathrm{EG}(2,2)$ code with $q=2$, $n=3$, $k=1$, $d=2$ has
  $w(X) = 1+X$. Here $\lambda_1 = d^2 = 4$ and
  (\ref{eq:equality_in_bound_circ}) holds in the form
\[
(1+X)^2 - X = 1+X+X^2
\]
so in this case $\lambda_2 = 1$ and $r(X) = 1$. Here
\[
\dmins = \wpsmin(\matr{H}) = n \left(
  \frac{2d-\lambda_2}{d^2-\lambda_2} \right) = 3 = q+1 \; .
\]
\end{example}
\vspace{-5mm}
\begin{example}\label{PG22example}
  The $\mathrm{PG}(2,2)$ code with $q=2$, $n=7$, $k=3$, $d=3$ has
  $w(X) = 1+X+X^3$. Here $\lambda_1 = d^2 = 9$ and
  (\ref{eq:equality_in_bound_circ}) holds in the form
\[
(1+X+X^3)(1+X^2+X^3) - 2X^3 = 1+X+\cdots+X^6
\]
so in this case $\lambda_2 = 2$ and $r(X) = 1$. Here
\[
\dmins = \wpsmin(\matr{H}) = n \left(
  \frac{2d-\lambda_2}{d^2-\lambda_2} \right) = 4 = q+2 \; .
\]
\end{example}
\vspace{-5mm}
\begin{example}\label{PG24example}
  The $\mathrm{PG}(2,4)$ code with $q=2$, $n=21$, $k=11$, $d=5$ has
  $w(X) = 1+X^2+X^7+X^8+X^{11}$. Here $\lambda_1 = d^2 = 25$ and
  (\ref{eq:equality_in_bound_circ}) holds in the form
\begin{eqnarray*}
& (1+X^2+X^7+X^8+X^{11})(1+X^3+X^4+X^9+X^{11}) \\ 
& - 4X^{11} = (1+X+X^2+\cdots+X^{20})(1-X+X^2)
\end{eqnarray*}
so in this case $\lambda_2 = 4$ and $r(X) = 1-X+X^2$. Here
\[
\dmins = \wpsmin(\matr{H})= n \left(
  \frac{2d-\lambda_2}{d^2-\lambda_2} \right) = 6 = q+2 \; .
\]
\end{example}
\vspace{-2mm}
Note that for a general $\mathrm{PG}(2,q)$ code, for the bound to hold
with equality we require
\begin{eqnarray*}
\wpsmin(\matr{H}) = q+1 & = & n \left( \frac{2d-\lambda_2}{d^2-\lambda_2}
 \right) \\ 
  & = & (q^2+q+1) \left( \frac{2(q+1)-\lambda_2}{(q+1)^2-\lambda_2} \right) \; .
\end{eqnarray*}
and therefore we must have $\lambda_2 = q$. Also, for a general
$\mathrm{EG}(2,q)$ code, for the bound to hold with equality we
require
\begin{eqnarray*}
\wpsmin(\matr{H}) = q+1 & = & n \left( \frac{2d-\lambda_2}{d^2-\lambda_2} 
\right) \\
& = & (q^2-1) \left( \frac{2q-\lambda_2}{q^2-\lambda_2} \right) \; .
\end{eqnarray*}
and therefore we must have $\lambda_2 = q$ if $q>2$, whereas for
$q=2$, any $\lambda_2$ will achieve the bound.

\section{Conclusions and Future Work}
\label{sec:conclusions:1}

A method has been presented for evaluation of the eigenvalue-based lower bound on the AWGNC pseudo-weight based on spectral
analysis, for QC and related codes. It was shown that the relevant eigenvalues may be found by computing the
eigenvalues of a certain number of small matrices. We also presented a 
necessary condition for the bound to be attained with
equality and gave a few examples of codes for which this happens.
Future work involves optimization of QC code designs based on these bounds. 

\section{Acknowledgment}
The first author was supported by NSF Grant DMS-0708033 and TF-0830608.
   
% \begin{enumerate}
% \item Use this method of finding eigenvalues to improve the bound of Tanner.
% \item Find conditions for the bound to hold.
% \item understand what the bound does exactly.
% \end{enumerate}
%%%%%%%%%%%%%%%%%%%%%%%%%%%%%%%%%%%%%%%%%%%%%%%%%%%%%%%%%%%%%%%%%%%%%%%%

% \bibliographystyle{ieeetr}
% \bibliography{../BibFiles-11-2006/huge,../BibFiles-11-2006/Pascal_ref,../BibFiles-11-2006/TF-proposal,../BibFiles-11-2006/bates,../BibFiles-11-2006/fuja} 
%\nocite{}
\end{document}